\documentstyle[osa,manuscript]{revtex}

\begin{document}
\title{The new scenario of the initial evolution of the Universe}
\author{V.Burdyuzha$^1$, O.Lalakulich$^2$, Yu.Ponomarev$^1$, G.Vereshkov$^2$}
\address{$^1$Astro Space Centre Lebedev Physical Institute \\
of Russian Academy of Sciences,\\
Profsoyuznaya 84/32, 117810 Moscow, Russia;\\
$^2$Scientific Research Institute of Physics\\
of Rostov State University\\
Stachki str. 194, 344104, Rostov on Don, Russia}
\maketitle

\begin{abstract}
We propose that the Universe created from "nothing" with relatively small
particles number and quickly relaxed to quasiequilibrium state at the Planck
parameters. The classic cosmological solution for this Universe with $%
\Lambda $-term has two branches divided by the gap. The quantum process of
tunneling between the cosmological solution branches and kinetic of the
second order relativistic phase transition in supersymmetric SU(5) model on
the GUT scale are investigated by numerical methods. Einstein equations was
solved together with the equations of relaxation kinetics. Other quantum
geometrodynamics process (the bounce from singularity) and the Wheeler- De
Witt equation are investigated also. For the formation of observable
particles number the model of the slowly swelling Universe in the result of
the multiple reproduction of cosmological cycles is arised naturally.
\end{abstract}

The inflation model and its basic modifications \cite{1} is very attractive
and explains many cosmological problems. Today it is the cosmological
paradigm \cite{2}. The standard conception of first order cosmological
relativistic phase transitions (RPT) from strong overcooled high symmetric
(HS) phase develops well \cite{3}. For the realization of this conception it
was necessary the inflaton potential with quite concrete properties (as rule
this potential must have a wide flat part). However, the problem of the
formation of observable particles number is not investigated in detail and
usually they speak that in the process reheating the energy of collisions of
bubble walls converts into energy individual quanta of scalar field, which
then decayed into normal particles. We discuss the case when the inflation
potential does not possess by specific properties, the Universe is created
from "nothing" and RPT is near to second order.

We proposed that:

I. Plasma and vacuum of the Universe, which was created from ''nothing'',
after processes of relaxation taking place near Planck parameters (see \cite
{4}), are in quasiequilibrium state. In our opinion the superearly Universe
created from ''nothing'' in anisotropic state (for example IX type of
Bianchi) with some number of particles and with some nonequilibrium state of
vacuum condensate.II. The topology of the Universe is closed. This
suggestion is caused by fact that only this Universe can be created from
''nothing'' (local properties of this Universe approach to local properties
of the flat Universe if the cosmological scenario solves the problem of the
flatness/entropy).III. After going out of the Universe from singularity
the initial number particles $N_o$ is large in comparison with unit ($%
N_o\sim 10^4\div 10^6$; but it is small in comparison with particles number
in observed Universe $(N_{obs.}\sim 10^{88})$.IV. RPT on the GUT scale ($%
T\sim 10^{16}$ Gev) which is more near to Planck scale is not the first
order RPT. This one is the second order RPT for which the generation of new
phase occurs by continuous mean \cite{5}.

One from possible series of RPT in the early Universe for the symmetry
breaking is: 
\begin{eqnarray*}
G &\Longrightarrow & [SU(5)]_{SUSY}\Longrightarrow [U(1)\times SU(2)\times
SU(3)]_{SUSY} \\
&\Longrightarrow & U(1)\times SU(2)\times SU(3)\Longrightarrow U(1)\times
SU(3)\Longrightarrow U(1)
\end{eqnarray*}

The only trace of first RPT is the initial $\Lambda$-term (the vacuum energy
density) connected with interactions of the local multidimentional
supergravity. The rest RPT are described by modern theories of elementary
particles. During RPT when cooling of cosmological plasma the vacuum
condensate with negative density of energy is produced. This condensate has
the asymptotic state equation $p_{vac} = - \epsilon_{vac} = const$. Thus the
RPT series are accompanied by generation of negative contributions in
cosmological $\Lambda$-term standing in Einstein equations. Accordingly to
observational data after all RPT the final $\Lambda$-term is zero
practically.For simplicity we have proposed the exact compensation of $%
\Lambda$-term already on the SUSY GUT energy scale. Our paper devotes the
quantitative model of the RPT $[SU(5)]_{SUSY} \Longrightarrow [U(1) \times
SU(2) \times SU(3)]_{SUSY}$ on the scale $\sim 10^{16}$ GeV. Some particles
acquare a rest mass which is proportional of an average value of Higgs field
after this gauge symmetry spontaneous breaking. The considered system
consists from third subsystems: 1) gas of massless particles, 2) gas of
massive particles interacting with vacuum condensate, 3) vacuum condensate.
The reactions of massless and massive particles on the cosmological
expansion is different. The change of these particles energy spectra because
of red shift happens on different laws. To get of evolution equations of
nonequilibrium system we use: a) the order parameter (OP) origin as C number
average of Higgs field; b) the method of getting of relaxation kinetics
equations for
subsystems particles analogous to described in \cite{6}; c) the method of
analysis of nonequilibrium relativistic systems analogous to described in 
\cite{7}; d) the estimation of creation particles local rate in variable OP
field which was got by the method analogous to described in \cite{8}.

The total system of equations of our theory involves: Einstein equations
with nonequilibrium energy momentum tensor of heterogeneity system

$\displaystyle \frac 3\kappa (\frac{\dot a^2}{a^2}+\frac 1{a^2})=\frac{k\pi
^2}{16}T^4+w(J_2+\eta ^2J_1)+\frac 1{2g^2}\dot \eta ^2+\frac 1{8g^2}(\eta
^2-m^2)^2,$%
\begin{equation}
\frac 1\kappa (2\frac{\ddot a}a+\frac{\dot a^2}{a^2}+\frac 1{a^2})=-\frac{%
k\pi ^2}{48}T^4-\frac w3J_2++\frac{\eta ^2(2J_1+\eta ^2J_o)}{3(4J_2+5\eta
^2J_1+\eta ^4J_o)}wD-
\end{equation}
$\hspace{1.0in}\displaystyle -\frac 1{2g^2}\dot \eta ^2+\frac 1{8g^2}(\eta
^2-m^2)^2$

the evolution equations for dissipative function $D$ and order parameter $%
\eta $

\pagebreak

$\displaystyle \dot D+(4\frac{\dot a}a+\frac 1\tau )D=$

\begin{equation}
\hspace{1.0in}=\frac{k\pi ^2T^4[(2J_1+\eta ^2J_o)(\eta ^2\frac{\dot a}a+\eta 
\dot \eta )+bT\dot \eta ^2]}{k\pi ^2T^4/4+w(4J_2+5\eta ^2J_1+\eta ^4J_o)}%
\left[ 1+\frac D{4J_2+5\eta ^2J_1+\eta ^4J_o}\right] 
\end{equation}

\begin{equation}
\ddot \eta +\left[ 3\frac{\dot a}a+bwg^2T\left( 1+\frac D{4J_2+5\eta
^2J_1+\eta ^4J_o}\right) \right] \dot \eta +
\end{equation}
$\hspace{1.0in}\displaystyle +\left\{ wg^2\left[ J_1+\frac{(2J_1+\eta ^2J_o)D%
}{4J_2+5\eta ^2J_1+\eta ^4J_o}\right] +\frac 12(\eta ^2-m^2)\right\} \eta =0$

and also the entropy equation, which can be transformed to the equation for
temperature of plasma: 
\begin{eqnarray}
\dot T+\frac{\dot a}aT=\frac{wT[(2J_1+\eta ^2J_o)(\eta ^2\frac{\dot a}a+\eta 
\dot \eta )+bT\dot \eta ^2]}{\frac{k\pi ^2T^4}4+w(4J_2+5\eta ^2J_1+\eta
^4J_o)}\left[ 1+\frac D{4J_2+5\eta ^2J_1+\eta ^4J_o}\right]
\end{eqnarray}
where $\kappa \simeq (10^{19}Gev)^{-2}$ is the Einstein constant; $k$ is the
number of bosons freedom degrees (exactly equal to the number of fermions
freedom degrees) the rest mass of which is equal to zero both in the high
symmetric (HS) phase and in the low symmetric (LS) phase; $w$ is an
analogous number of particles freedom degrees acquaring of mass $\eta $ in
LS phase; $g$ is the gauge coupling constant of particles with vacuum
condensate; $m=const$ is the limited value of particles mass when $%
T\rightarrow 0$ in the LS phase; $b$ is the numerical coefficient unity
order; $\tau $ is the time of relaxation between subsystems massless and
massive particles (we use units in which $\hbar =c=1$);

$\displaystyle J_n(T,\eta )=\frac 1{2\pi ^2}\int_o^\infty \frac{dpp^{2n}}%
\omega (\frac 1{\exp \frac \omega T-1}+\frac 1{\exp \frac \omega T+1})$

$\displaystyle n=0,1,2$ \hspace{3mm}$\omega =(p^2+\eta ^2)^{\frac 12}$

characteristic integrals through which observable magnitudes of our model
are expressed.In the theory (1)-(4) the property of the early Universe
depends on total number of particles in closed space the critical value of
which is: $\displaystyle N_{cr}\equiv 7\xi (3)(\frac{k+w}2)^{1/4}(\frac{12g}{%
\pi \kappa m^2})^{3/2}\simeq 5\times 10^{11},$

where $\xi (3)$ is the zeta function of Riemann. If the value of initial
total particles number in HS plasma $N_o<N_{cr}$ then the cosmological
solution contains two branches divided by the gap (Fig. 1). The branch I is
the Friedmann solution which is distorted slightly by $\Lambda $-term and
the branch II is the de Sitter solution which is distorted slightly by
matter. The investigation of the plasma evolution temperature regime shows
that on the branch I the minimal temperature $T_{I(min)}$ (corresponding to
max radius $a_{I(max)}$ is essentially more than critical temperature $T_{c}=
(\frac{4m^2}{wg^2})^{1/2}$ defining the boundary of thermodynamical
stability of the HS phase. On the branch II the maximal temperature $
T_{II(max)}$ (corresponding of min radius $a_{II(min)}$ is essentially less
than $T_c$. From this hierarchy of temperatures it follows that: 1) the
Universe evoluting on the branch I doesn't undergo to RPT in the LS phase
(it goes out from initial singularity and come to final one in the HS phase;
2) on the branch II the Universe cann't be in the HS phase in principle. The
branch II is classically prohibited. There are two reasons for it: at first
the branch II is separated from the branch I by classically nonovercome gap;
at second the branch II is thermodynamicaly instable. Note that the branch
II exists formally when nonzero $\Lambda $-term takes place. When plasma and
vacuum are in the LS phase and $\Lambda $-term is equal to zero, the branch
II doesn't appear at all. From the classical point of view the branch II
does not exit, i.e. in general it exists only virtualy as classical
unrealized variant of evolution. Therefore if the Universe, which was
created from ''nothing'', has particles number less than critical one, then
this Universe can not transfer in the macroscopic object containing observed
particles number.

But the situation is changed radically in quantum geometrodynamics (QGD) of
the closed Universe. In this theory there is a small but nonzero probability
of tunneling through the gap dividing the branch I and the branch II of the
classical solution. Let us to discuss shortly of the mathematical model of
this phenomenon.

The problem is to construct the quantum analogue of equations (1)-(4) i.e.
the equation of Wheeler-De Witt (WDW) for the Universe wave function $\psi
=\psi (a,\eta )$. However, the dissipative dynamics of system (1)-(4) isn't
Hamilton one since formal methods of quantization don't apply to solution of
this problem. The necessary of the new quantum theory, which must correlate
with the second law of thermodynamics, was discussed by Penrose \cite{9}.
The absent of this theory compells us to decide of this task in two steps.
On the first stage the dissipative processes aren't take into account. We
hope the quantum nondissipative geometrodynamics reflects approximately the
properties of processes of tunneling through barrier and bounce from
singularity. On the second stage the dissipative processes are described by
the classical method on basic of the eqs. (1)-(4).

The quantization nondissipative system is the realization of the
Lapchinsky-Rubakov's idea \cite{10} suggested, to describe the present of
matter in the closed Universe by methods of the effective potential involved
in WDW equation. The fact of the existence of Hamiltonian bond is needed in
another discussion. This bond is caused by the gauge invariance of the
theory relatively the time transformation. The arbitrary in time choice is
removed by an addition condition which is put on the gauge variable. This
variable is the algebraic form of metric components containing the $g_{oo}$%
-component. For example, if the gauge variable is $g_{oo}=\lambda ^2$ then
the WDW equation doesn't depend from choice of the additional condition $%
F(\lambda ,a,\eta )=0$. However, we can bring the local - conform
transformation of time and gauge variable 
\begin{eqnarray}
dt=af(a)dt^{^{\prime }},\hspace{2mm}\lambda =\lambda ^{^{\prime }}af(a),
\end{eqnarray}
where $f(a)$ is an arbitrary function. In this case the equation doesn't
depend on the gauge condition $F(\lambda ^{^{\prime }},a,\eta )=0$ but it
depends on the generator of local-conform transformation $f(a)$. Formally
from the mathematically point of view this dependence is caused by the
nonlinear coupling of gauge variable $\lambda =\lambda ^{^{\prime }}af(a)$
with square of generalized impulse $p\sim \dot a$. The every variant putting
order of operators $p$ and $f(a)$ which coincide with the property of
hermiticity of Hamiltonian generates additional members in the WDW equation.
These members can be interpretated as additional contribution in potential
of WDW equation: 
\begin{eqnarray}
U_f(a)=\frac \kappa {24\pi ^2}\left[ \frac 1{4f}\frac{d^2f}{da^2}-\frac 3{%
16f^2}(\frac{df}{da})^2\right] .
\end{eqnarray}
Thus the effect of the spontaneous breaking of symmetry relatively
locally-conform time transformations (5) takes place in the WDW theory for
the closed isotropic Universe. The additional contribution (6) has the sense
of energy of some gravitational vacuum condensate (GVC) the production of
which fixes the breaking of discussed symmetry in whole space of the closed
Universe. We propose the GVC must secure the bounce from singularity i.e.
prolong the time of the Universe existence. This result will take place if 
\begin{eqnarray}
f(a)=a^{4(S+1)},\;\;U_{f(a)}\equiv U_{S(a)}=\frac \kappa {24\pi ^2}\frac{%
S(S+1)}{a^2},
\end{eqnarray}
where $S=const>0$ is the parameter of the GVC. The WDW equation in this case
is written for the Universe wave function $\psi =\psi _{NS}(a,\eta )$
depending on two variables $\;a,\eta $ and on two parameters $N$ and $S$

$\displaystyle -\frac \kappa {24\pi ^2}\frac{\partial ^2\psi _{NS}}{\partial
a^2}+\frac{g^2}{4\pi ^2a^2}\frac{\partial ^2\psi _{NS}}{\partial \eta ^2}+$%
\begin{equation}
\hspace{1.0in}+\left[ \frac \kappa {24\pi ^2}\frac{S(S+1)}{a^2}+\frac{6\pi ^2%
}\kappa a^2-2\pi ^2a^4\epsilon _N(a,\eta )-\frac{\pi ^2}{4g^2}a^4(\eta
^2-m^2)^2\right] \psi _{NS}=0,
\end{equation}
where $\epsilon _N(a,\eta )$ is the energy density of subsystems of
particles the mathematics form for which coincides with thermodynamical
expression. As we have proved the equation (8) has the most important
property: the solution located on the HS vacuum (near $\eta =0$) for every
values of the Universe radius is among its solutions. The main dependence of
a wave function quasilocated on the HS vacuum on the Universe radius can be
factorired by the separate function. This function satisfies the equation
which is formally similar to the Schrodinger one for stationar states of
some conditional ''particle'' in potential field $U(a)$ the shape has been
shown on Fig. 2. The asymptotics of the potential $U(a)$ in the small $a$
region is defined by the GVC energy (7). From Fig. 2 one can see that the
GVC provides the quantum bounce from singularity. The quantum bounce
hypothesis being accepted leads to this that Universe with small particles
number oscilates quasiclassically in the region I of potential $U(a)$. The
sector II of potential $U(a)$ corresponds to the branch II of the classical
cosmological solution (compare Fig. 1 and Fig.2). The probability of
tunneling through the barrier dividing the branches I and II is exponential
small when a particles number is small. It is increased monotonuously with
increase the number of oscillation cycles in the region I. If the number of
oscillation cycles is large then the causely-consequence connections have
been set up among all its space-time points and in this case the problem of
horizon is absent. We propose that after tunneling the causely-consequence
connections among different points of the Universe are conserved.

Thus our Universe was created from "nothing" with small particles number and
performed exponentially long oscillations in the region I of effective
potential $U(a)$. Here the Universe existed in the HS phase. After large
number of oscillations the Universe has undergone to the tunneling
transition in the HS thermodynamically instable phase. If after tunneling
the Universe is appeared directly near the barrier i.e. on the left boundary
of the region II then it size is increased multiplely: 
\begin{eqnarray}
\frac{a_{II(min)}}{a_{I(max)}} = 2(\frac{N_{cr}}{N_{o}})^{2/3}
\end{eqnarray}
Accordingly to (9) when the initial particles number $N_{o} = 5 \times
10^{5} $ radius increases in $2 \times 10^{4}$ times in the result of the
quantum tunneling. This phenomenon can be considered as analogue of
classical inflation. After tunneling the Universe will be found in strongly
nonequilibrium state. The relaxation of nonequilibrium plasma and vacuum to
new equilibrium state corresponding the stable of LS phase is coming in the
relaxation kinetics regime and is accompanied by sharply entropy increase.
This process is described by the eqs. (1)-(4).

The classical state corresponding to minimum curve $a_{II}$ must appear with
the greatest probability. The evolution of the Universe with the initial
state $a(t_{o}) = a_{II(min)}$ and with $N_{o} \simeq 5 \times 10^{5}$ shows
on Fig. 3-5. The RPT begins in fact immediately after tunneling and is
accompanied by nonlinear OP vibrations wit the frequency $\sim T_{c}$ (Fig.
3) and particles creation (Fig. 4). After damping of these vibrations and
the RPT finish the particles number in the Universe is increased in $2.6
\times 10^{6}$ times (for $N_{o} = 5 \times 10^{5}$). The numerical
experiments have shown that the total particles number in the Universe after
relaxation coincides approximately with $N_{cr}$ for different values $N_{o}
\sim 10^{2} \div 10^{9}$. Thus he particles number in the macroscopic
Universe after tunneling and RPT is expressed through the fundamental
constants:

\begin{eqnarray}
N_{cr}\sim \kappa ^{-3/2}m^{-3}=(\frac{10^{19}Gev}m)^3
\end{eqnarray}

The classical evolution of the closed Universe shown on Fig. 3-5 finishs in
singularity. However the calculation of QGD effects transfers the enter in
singularity on the quantum bounce at the Planck parameters. The number of
particles is conserved during this bounce since next classical evolution
cycle starts for $N > N_{cr}$.The numerical research of this Universe
evolution cycle was performed also. The main result is the conclusion that
in the Universe with $N > N_{cr}$ the second order RPT happens
quasiadiabatically. Relative increase of particles number for total
evolution cycle is smaller than one percent. Thus we got the model of he
slowly swelling Universe.

Further detalization of the scenario of the Universe evolution requires the
calculation of RPT on others energetic scales. The Universe during of
evolution must overcome some potential barriers similar to the barrier which
is shown on Fig. 2. Accordingly (10) after last RPT on scale $\sim 100 Mev$
we obtain the macroscopic Universe with particles number $N\sim 10^{60}$.
The problem is that $N\sim 10^{88}$ in the observable Universe. We can
formulate some hypothesis explaning as to obtain this. 1. The observable
particles number has been created after multiple reproduction of
cosmological cycles containing all series of RPT. During every cycle the
particles number is increased in comparison with a previous one because of
dissipative processes accompaning of RPT. 2. After tunneling through some
barrier on some cycles of evolution the Universe can appear in a point of
trajectory which is sufficiently far from left boundary of the region II
(see Fig. 2). From the solution of the WDW equations it follows that the
probability of this process is decreased exponentially as far as moving away
from barrier. The numerical experiments, have shows that particles number
creating during of RPT which is delayed on time, increases exponentially
with time delay. 3. Effects of strong nonlinear interaction of different
vacuum condensates in more complex series of RPT lead to time delay of these
RPT (as example, the series $G\Longrightarrow E_6\Longrightarrow
O(10)\Longrightarrow SU(5).....$ can be considered). 4. The fourth variant
connects with the hypothetical possibility of dynamics chaos regime in the
region of nonequilibrium RPT.\\

\begin{figure}[tbp]
\caption{The cosmological solution for closed Universe for initial particles
number $N_{o} \ll N_{cr}$. The units of time and scale factor are: $t_{o}
\simeq 3.5 \times 10^{-38}$ sec, $a_{o} \simeq 1.5 \times 10^{-27}$ cm.}
\end{figure}
. 
\begin{figure}[tbp]
\caption{The dependence of the Universe wave function factorized by the
separate function $U(a)$ from scale factor $a$.}
\end{figure}
\begin{figure}[tbp]
\caption{The change of order parameter ($\eta$) during the Universe
evolution. The units of time and scale factor are the same as Fig.1.}
\end{figure}
\begin{figure}[tbp]
\caption{The change of relative particles number ($N/N_{o}$) during the
Universe evolution. The units of time and scale factor are the same as
Fig.1. }
\end{figure}
\begin{figure}[tbp]
\caption{The change of scale factor {\it (a)} during the Universe evolution.
The units of time and scale factor are the same as Fig.1.}
\end{figure}


\begin{references}
\bibitem{1}  A.H.Guth. Phys Rev. {\bf D 23}, 347 (1981); K.Sato. Mon. Not.
R. Astron. Soc. {\bf 195},467 (1981); A.D.Linde. Phys. Lett. {\bf 108B}, 389
(1982; A.Albrecht and P.J.Steinhardt. Phys. Rev. Lett. {\bf 48}, 1220
(1982); A.D.Linde. Phys. Lett. {\bf 129B}, 177 (1983); A.A.Starobinsky. JETP
Lett. {\bf 42}, 152 (1985); J.Silk and M.Turner. Phys. Rev. {\bf D35}, 419
(1986); A.Dobado and A.L.Maroto. Phys. Rev. {\bf D52}, 1895 (1995);
M.Bucher, A.S.Goldhaber, N.Turok. Phys. Rev. {\bf D52}, 3314 (1995); B.Ratra
and P.J.E.Peebles. Phys. Rev. {\bf D52}, 1837 (1995); M.White and D.Scott.
Astrophys. J. {\bf 459}, 415 (1996).

\bibitem{2}  M.S.Turner. In Proceedings of Texas Symposium in Relativistic
Astrophysics, p. 153 (1995).

\bibitem{3}  E.W.Kolb, in Proceedings conf. ''Evolution of the Universe and
its Observational Quest'', p. 31 (1994).

\bibitem{4}  G.M.Vereshkov et al. JETP {\bf 73}, 1985 (1977); Y.Shtanov,
J.Traschen, R.Brandenberger. Phys. Rev. {\bf D 51}, 5438 (1995);
G.M.Vereshkov, A.N.Poltavtsev. JETP, {\bf 71}, 3 (1976)

\bibitem{5}  V.V.Burdyuzha, O.D.Lalaculich, Yu.N.Ponomarev, G.M.Vereshkov
Intern. J. of Modern Phys. D., 5, N 2, 1996, Astro-ph-9604124

\bibitem{6}  Ya.B.Zeldovich, Yu.P.Raizer. Physics of shock waves and high
temperature hydrodynamic phenomena, in Chapter VIII, volume 2, Academic
Press, 1967.

\bibitem{7}  L.D.Landau, E.M.Livshits. The hydrodynamics, in Chapter 15,
Nauka (1986).

\bibitem{8}  Ya.B.Zeldovich, A.A.Starobinsky, JETP Lett. {\bf 26}, 373
(1977).

\bibitem{9}  R.Penrose. The poster talk on Texas Symp. on Relativistic
Astrophysics, Munchen, 1994.

\bibitem{10}  V.G.Lapchinsky, V.A.Rubakov. Acta Phys. Polonica {\bf B10},
1041 (1979).
\end{references}
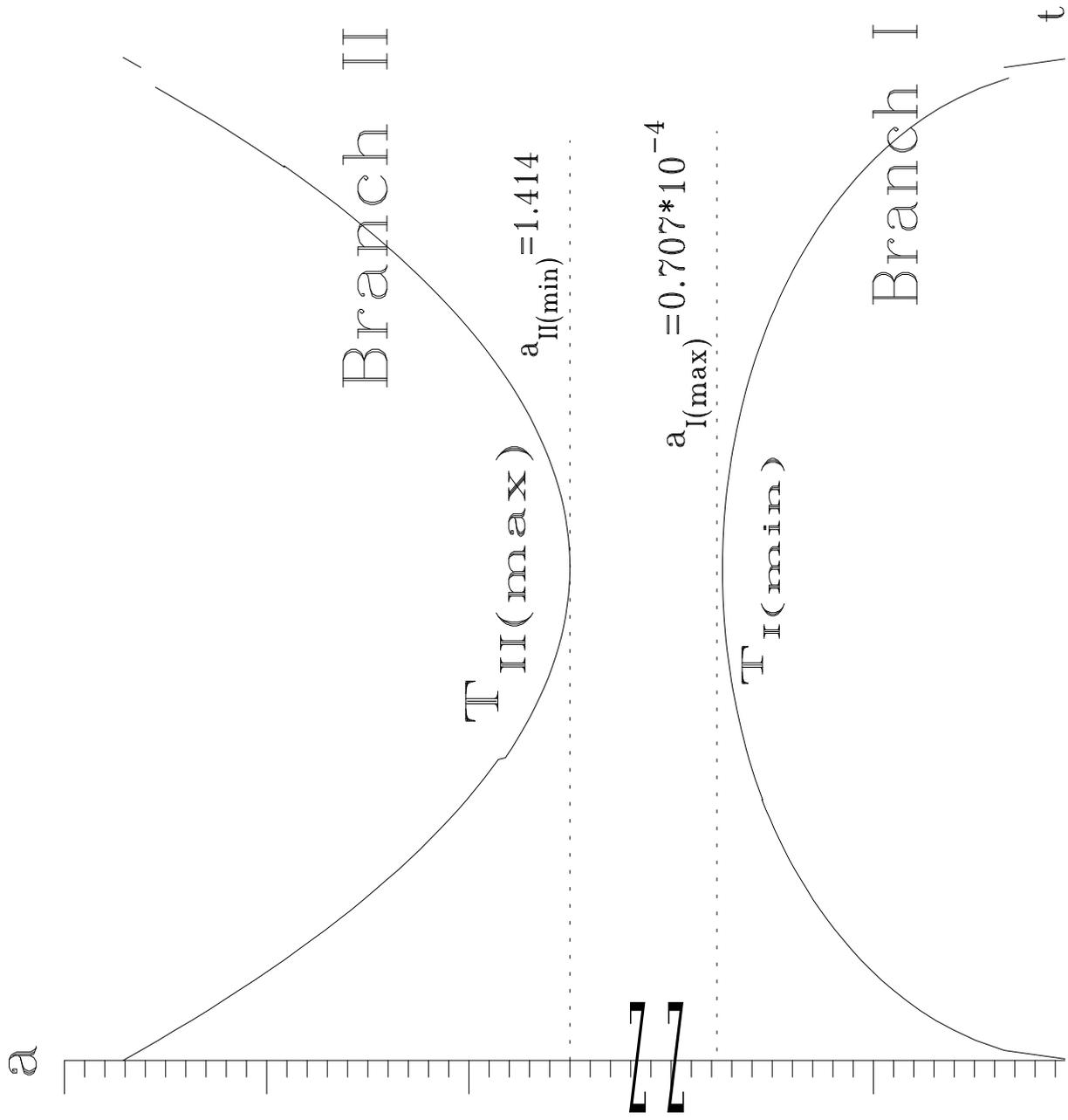
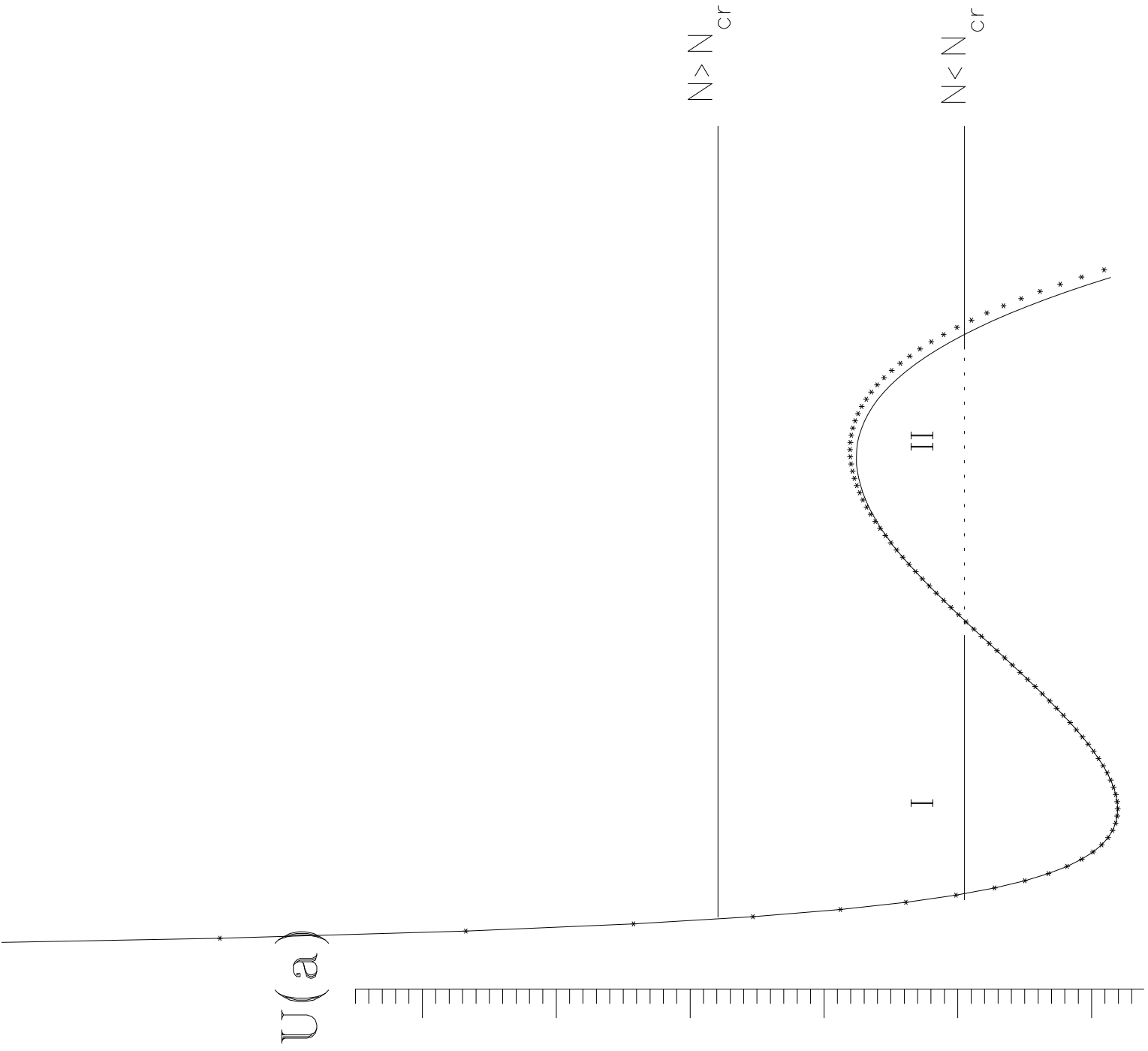
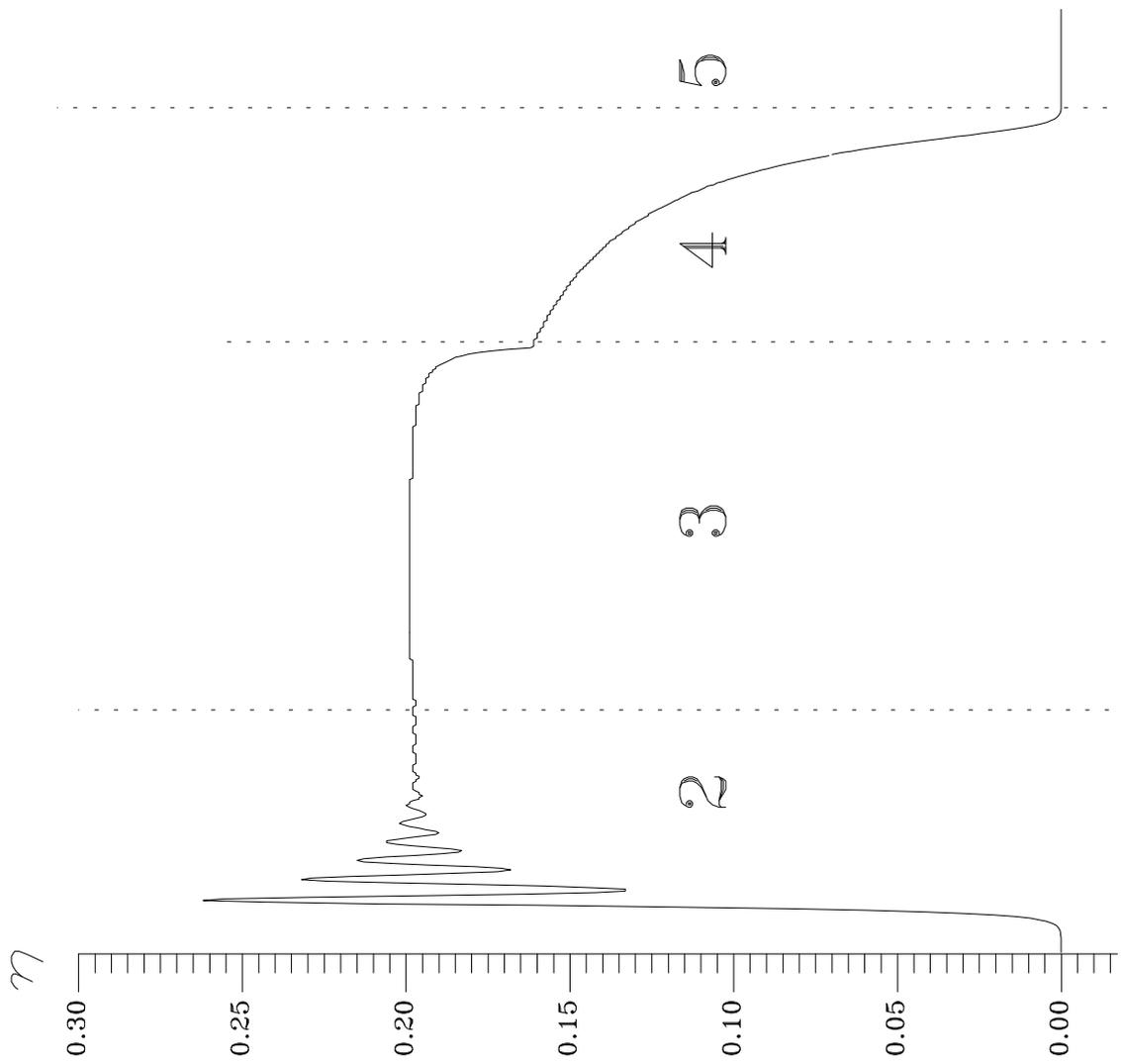
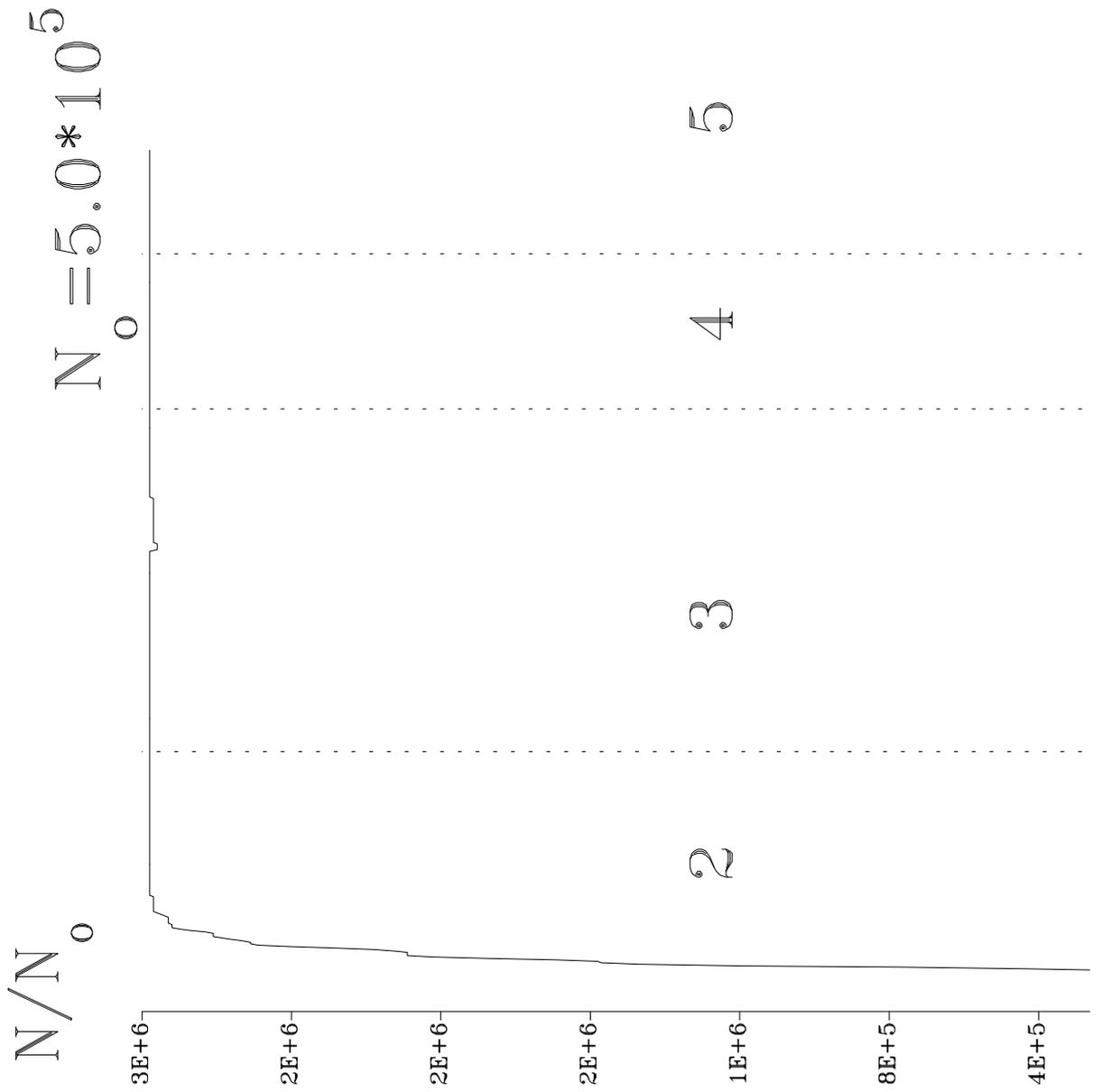
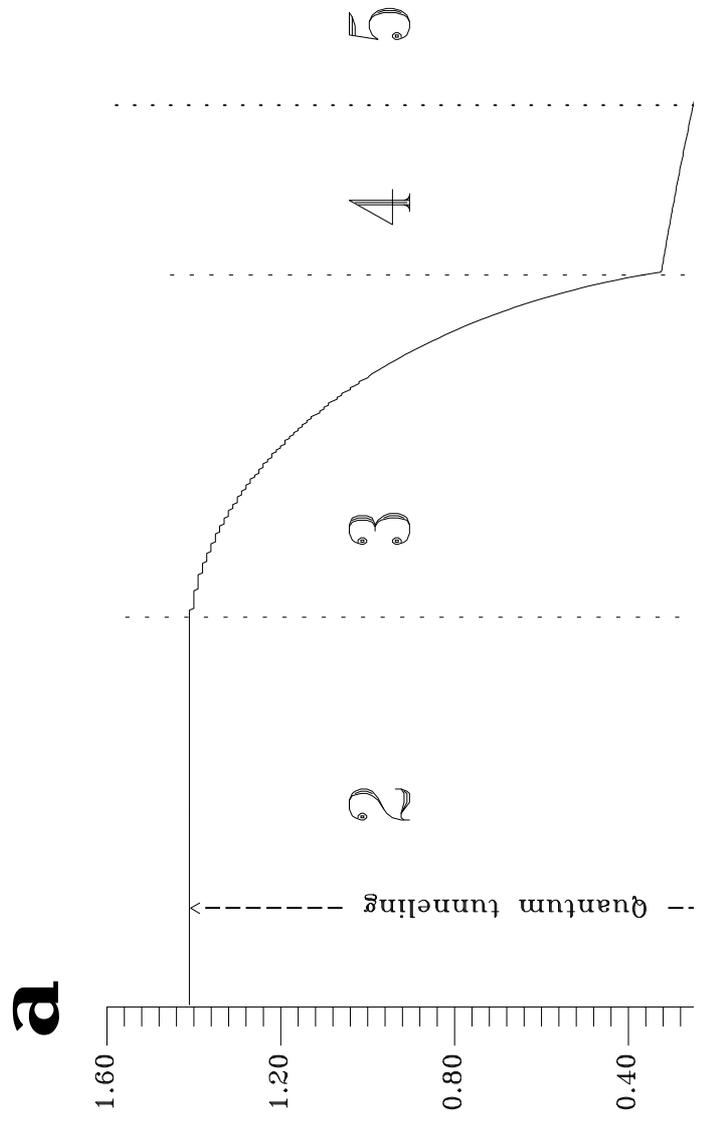
\end{document}